\documentclass[%
 reprint,
 amsmath,amssymb,
 aps,
]{revtex4-2}

\usepackage{graphicx}
\usepackage{dcolumn}
\usepackage{bm}

\usepackage{xcolor}

\newcommand{\Vmat}{{\boldmath $\cal V$}}
\newcommand{\Kmat}{{\boldmath $\cal K$}}

\newcommand{\numat}{\mbox{\boldmath $\nu$}}

\newcommand{\Smat}{\mbox{\boldmath $\cal S$}}

\newcommand{\Xmat}{\mbox{\boldmath $X$}}

\begin{document}

\preprint{APS/123-QED}

\title{Dissociative recombination of NeH$^{+}$ with low-energy electrons: Multichannel quantum defect theory including non-adiabatic couplings}

\author{Riyad Hassaine\textsuperscript{1}, J\'anos Zsolt Mezei\textsuperscript{2}, Dahbia Talbi\textsuperscript{3}, Jonathan Tennyson\textsuperscript{4,1}, and Ioan F. Schneider\textsuperscript{1}}
\affiliation{\textsuperscript{1}Laboratoire Ondes et Milieux Complexes, CNRS, Université Le Havre Normandie, 53 rue Prony, Le Havre, 76058, France}
\affiliation{\textsuperscript{2}HUN-REN Institute for Nuclear Research (ATOMKI), H-4001 Debrecen, Hungary}
\affiliation{\textsuperscript{3}LUPM CNRS-UMR5299, Universit\'e de Montpellier, F-34095 Montpellier, France}
\affiliation{\textsuperscript{4}Department of Physics and Astronomy, University College London, Gower Street, London WC1E 6BT, United Kingdom}

\date{\today}

\begin{abstract}
A theoretical investigation of the dissociative recombination (DR) of NeH$^{+}$ with low-energy electrons in the regime where the process occurs without direct potential energy curve crossings is presented. The calculations are performed using multichannel quantum defect theory, incorporating non-adiabatic couplings between electronic states. Unlike the previous treatment of the DR of HeH$^{+}$, where only first-order radial couplings $A(R)$ were considered, our formulation also incorporates the second-order terms $B(R)$, together with a radial density of states $\beta_\nu(R)$ to describe the transition into the ionization continuum. This development uses a large number of potential energy curves and non-adiabatic couplings of NeH characterized by us previously, enabling a consistent modeling of the DR process. The resulting cross sections show good agreement with the available experimental data and fill a gap in theoretical data below 4.5~eV, where no detailed quantum calculations are currently available. For $v_i^{+}=0$, the isotropic thermal DR rate coefficient of NeH$^{+}$ spans $5.2\times10^{-8}$~cm$^{3}$\,s$^{-1}$ at $T=10$~K to $6.1\times10^{-9}$~cm$^{3}$\,s$^{-1}$ at $T=4000$~K. Over the same range, experimental values for ArH$^{+}$ are one to two orders of
magnitude smaller, while astrophysical models so far have assumed a constant value of
$10^{-11}$~cm$^{3}$\,s$^{-1}$ for both ArH$^{+}$ and NeH$^{+}$. Our results on NeH$^{+}$ could account for the non-detection of NeH$^{+}$ in nova and supernova remnants, due to abundant depletion from efficient dissociative recombination. In fusion edge plasmas containing neon, NeH$^{+}$ can form as a transient impurity through reactions with vibrationally excited hydrogen molecular ions. Our anisotropic DR rate coefficients for $v_i^{+}=0$--2 fall in the $10^{-9}$--$10^{-8}$~cm$^{3}$\,s$^{-1}$ range (e.g.\ at 0.1 and 2.0~eV they are $4.4\times10^{-9}$ and $1.2\times10^{-9}$ for $v_i^{+}=0$, 
$2.7\times10^{-8}$ and $3.9\times10^{-9}$ for $v_i^{+}=1$, and $4.9\times10^{-8}$ and $6.7\times10^{-9}$~cm$^{3}$\,s$^{-1}$ for $v_i^{+}=2$). These values are not negligible for molecular ions and indicate that DR efficiently converts NeH$^{+}$ back into neutral neon and hydrogen, preventing significant accumulation in the divertor region.

\end{abstract} 

\maketitle

\section{Introduction}
Neon is actively being considered as a radiative coolant in the divertor region of the International Thermonuclear Experimental Reactor (ITER)~\cite{zajfman1996, mitchell2001, mitchell2005, ngassam2008} due to its ability to dissipate thermal energy via electron-impact excitation and ionization processes. Following ionization, neon ions can undergo dielectronic recombination with electrons and subsequent radiative decay,
\begin{equation}
\text{Ne}^+ + e^- \rightarrow \text{Ne}^{**} \rightarrow \text{Ne} + h\nu,
\end{equation}
providing an efficient mechanism for energy removal through photon emission.

In the edge plasma, where vibrationally excited H$_2^+$ ions are abundant, collisions with neutral neon atoms may lead to the formation of NeH$^+$:
\begin{equation}
\text{Ne} + \text{H}_2^+(v_i^+ > 1) \rightarrow \text{NeH}^+ + \text{H},
\end{equation}
a process that becomes exothermic only above the first vibrational level of  H$_2^+$~\cite{mitchell2005}. The presence of NeH$^+$ in this region calls for a detailed understanding of its reactivity, particularly in its interactions with electrons.

Among electron-driven processes, dissociative recombination (DR) and vibrational transitions (VT) are of special relevance:
\begin{equation}
\text{NeH}^+ + e^- \rightarrow \text{NeH}^*, \text{NeH}^{**} \rightarrow
\begin{cases}
\text{Ne} + \text{H} & \text{(DR)} \\
\text{NeH}^+ + e^- & \text{(VT)}
\end{cases}.
\label{DRVT}
\end{equation}
DR cross sections for NeH$^+$ were measured at the ASTRID storage ring in 2005 by Mitchell \textit{et al.}~\cite{mitchell2005}, and theoretical studies have addressed the 4.5–22~eV collision energy range~\cite{ngassam2008}, with reasonable agreement between 6 and 10~eV. However, no theoretical predictions were available for electron collision energies below 4.5~eV, where experimental data reveal cross sections between $10^{-21}$ and $10^{-15}$~cm$^2$~\cite{mitchell2005}. This lower energy region is more important for practical applications in both fusion plasmas and astrophysics.

HeH$^+$, believed to be the first molecule formed in the Universe, was observed in 2016 in the planetary nebula NGC7027~\cite{gusten2019}. As the simplest heteronuclear molecular cation (since HD$^+$ or HT$^+$ is usually treated as H$_2^+$), it serves as a prototype for indirect dissociative recombination, proceeding exclusively through electron capture into Rydberg states. This process occurs in the absence of any neutral diabatic state crossing the ionic potential energy curve~\cite{jt152,guberman1994}. State-selective measurements performed at the Cryogenic Storage Ring (CSR)~\cite{novotny2019} have shown a significant decrease in DR rate coefficient at very low electron collision energies, a feature that may contribute to enhanced HeH$^+$ abundances in cold astrophysical environments. These findings have been supported by recent multichannel quantum defect theory calculations employing ro-vibrational frame transformation techniques~\cite{vcurik2020a,hvizdovs2020b}.

A similar situation arises for ArH$^+$. Before its detection in the Crab Nebula~\cite{barlow2013}, experimental measurements by Mitchell \textit{et al.}~\cite{mitchell2005b} using the ASTRID storage ring reported DR rate coefficients below $10^{-9}$~cm$^3$s$^{-1}$ for electron energies below 2~eV. This low reactivity was understood in light of the high asymptotic energy (1.822~eV) of the lowest diabatic neutral state~\cite{abdoulanziz2018,djuissi2022}, which — under the assumption of purely Rydberg-valence coupling — implied that recombination was unlikely to occur below this threshold. However, recent measurements by Kalosi \textit{et al.}~\cite{kalosi2024} revealed low but non-negligible DR rates into the ground state, attributed to non-adiabatic couplings between the ionization continuum and the ground electronic state. This scenario is supported by theoretical considerations in the same study, as well as computations in progress by Larson and Orel~\cite{kalosi2024}.

Unlike HeH$^+$ and ArH$^+$, NeH$^+$ has not yet been detected in any astrophysical environments. Recent astrochemical models suggest that NeH$^{+}$ could be efficiently formed in nova remnants through
\[
\text{Ne} + \text{HeH}^+ \rightarrow \text{NeH}^+ + \text{He},
\]
with a rate coefficient of approximately $10^{-9}$~cm$^3\cdot$s$^{-1}$~\cite{sil2024}, a reaction more favorable than the alternative
\[
\text{Ne}^+ + \text{H}_2 \rightarrow \text{NeH}^+ + \text{H},
\]
proposed by Theis \textit{et al.}~\cite{theis2015}. In stellar environments, the abundance of neon is much higher than that of argon~\cite{schwarz2002,Das2020}, which in principle favors NeH$^{+}$ over ArH$^{+}$. Nevertheless, the overall formation of NeH$^{+}$ remains limited by the low cosmological abundance of neon and, in regions where neon is ionized, by the relative scarcity of molecular hydrogen. In addition, a high DR rate at low temperatures would significantly reduce its equilibrium abundance, which may explain its current non-detection.

To better understand the possible role of dissociative recombination in limiting the abundance of NeH$^+$, it is necessary to develop a detailed theoretical description of the process. This requires accurate molecular data, including potential energy curves and non-adiabatic couplings between the relevant electronic states of the neutral species which we provided in a recent study~\cite{hassaine2025}.

The present work reports the full theoretical treatment of the low-energy DR of NeH$^+$, by incorporating non-adiabatic couplings between relevant neutral states and quantum defects associated to the highest computed Rydberg state for both symmetries, into the multichannel quantum defect theory formalism. This enables the computation of DR cross sections in the low-energy regime most relevant to fusion-edge and to the plasma conditions found in stellar remnants.

\section{Theoretical method}

\begin{figure}[t]
\centering
		\includegraphics[width=0.90\columnwidth]{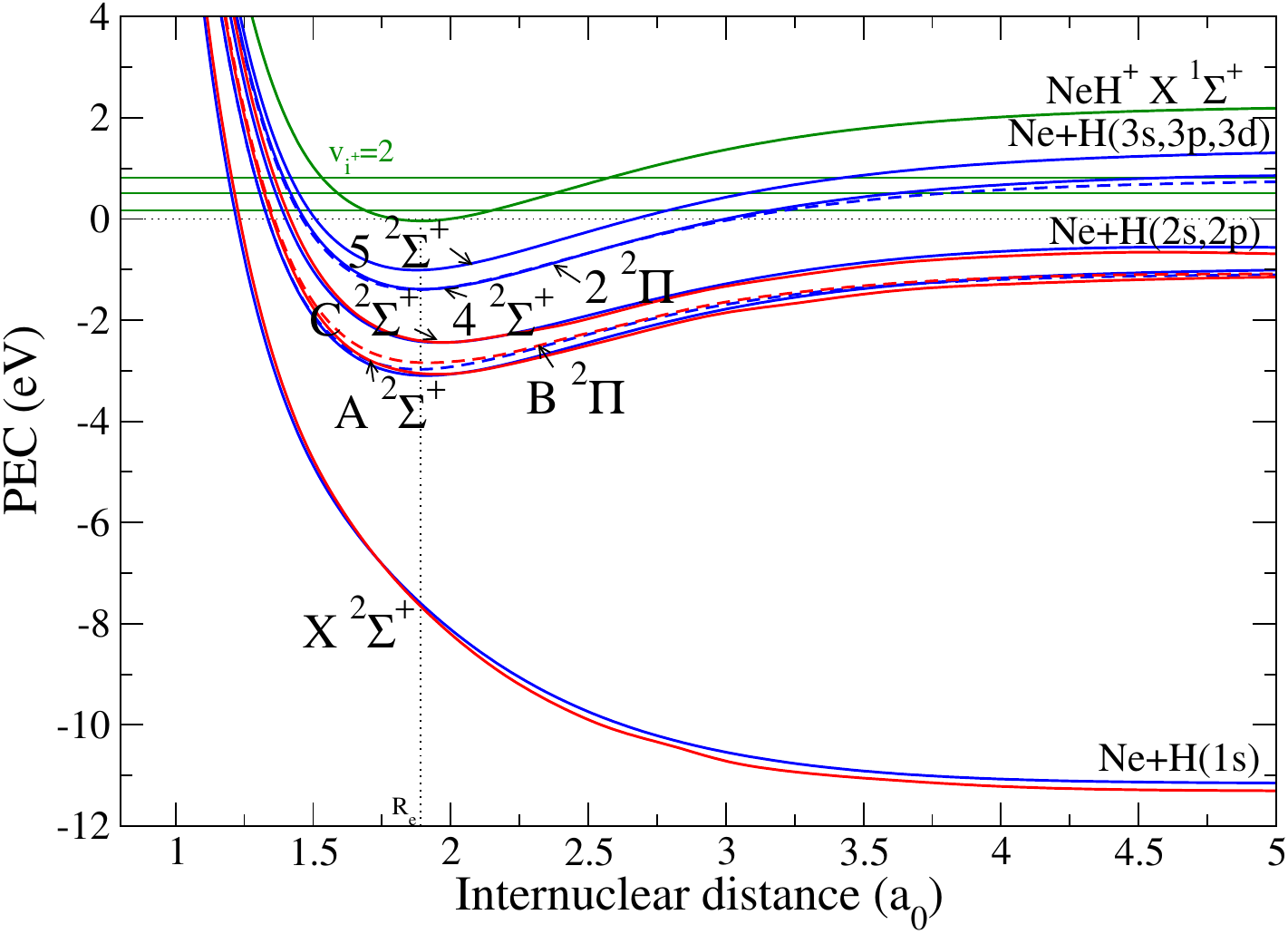}
    \caption{\textit{Ab initio} potential energy curves (PECs) of the ground electronic state of NeH$^{+}$ and of the lowest electronic states of neutral NeH—both ground and Rydberg—compiled from~\cite{hassaine2025}. Blue solid lines represent neutral states of $^2\Sigma^{+}$ symmetry, blue dashed lines correspond to $^2\Pi$ symmetry, and the green solid line denotes the ionic ground state X~$^1\Sigma^{+}$. The first three vibrational levels of NeH$^{+}$, $v_i^+ = 0$, 1, and 2, are also shown in green. Our results are compared with the calculations of Theodorakopoulos \textit{et al.}~\cite{Theodorakopoulos1987} (red lines).}
    \label{fig:1}
\end{figure}
\begin{figure}[t]
\centering
		\includegraphics[width=0.90\columnwidth]{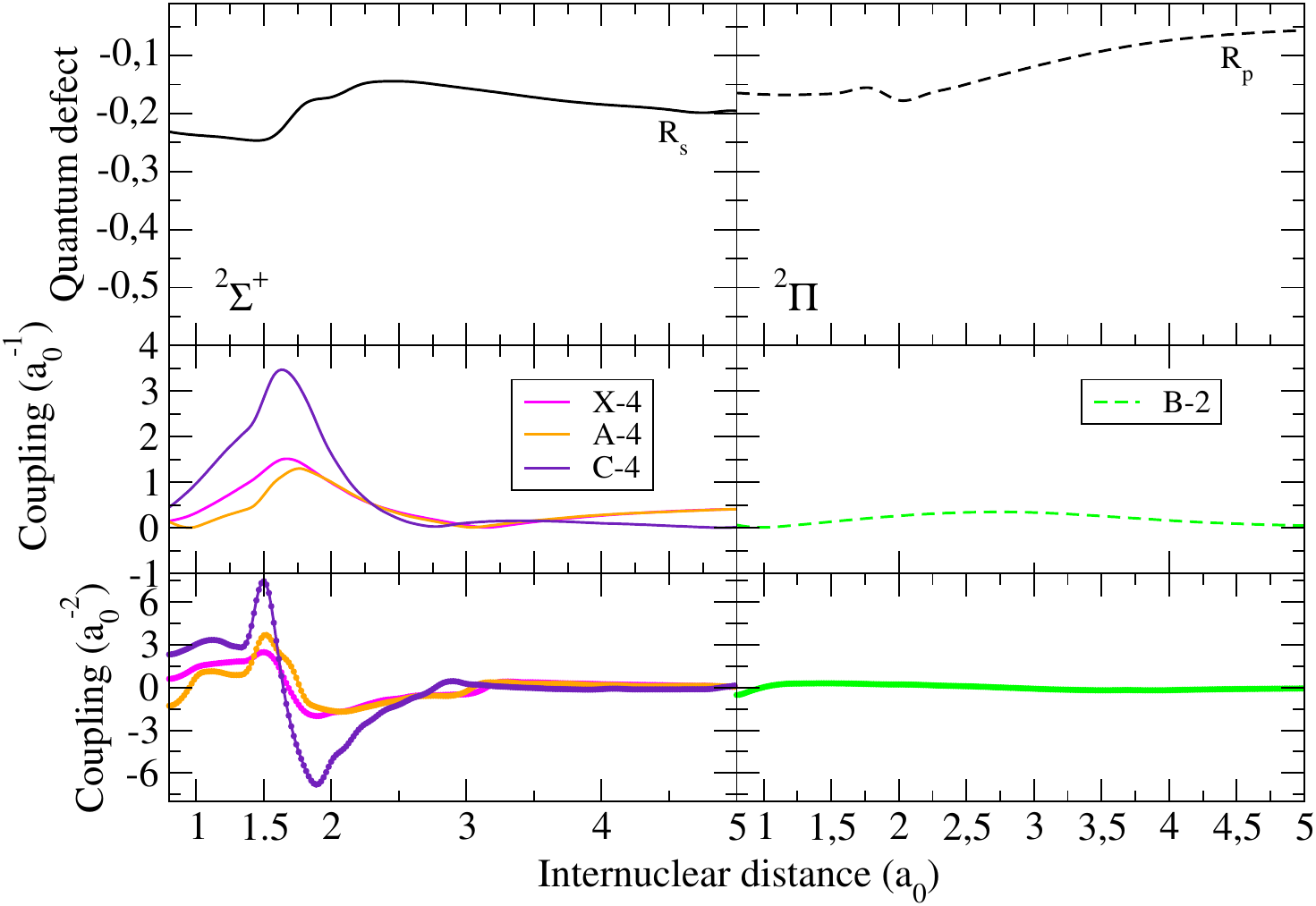}
    \caption{Relevant molecular data used in the nuclear dynamics calculations. The quantum defects $\mu_{\ell}^{\Gamma}(R)$ characterizing the Rydberg series of partial waves—$s$ for the $^2\Sigma^{+}$ symmetry (left panel) and $p$ for the $^2\Pi$ symmetry (right panel)—are shown in the top row. In the middle and bottom rows, the orange, pink, and purple solid curves represent the non-adiabatic couplings $A(R)$ and $B(R)$ for the $^2\Sigma^{+}$ symmetry (left panel), while the green curves correspond to the same couplings for the $^2\Pi$ symmetry (right panel).}
    \label{fig:2}
\end{figure}

Our recent study~\cite{hassaine2025} established the molecular structure foundations for modeling the DR of NeH$^+$, by computing accurate adiabatic potential energy curves (PECs) and non-adiabatic couplings (NACs) between relevant neutral states; these are summarized in Figs.~\ref{fig:1} and \ref{fig:2}. The potential energy curve calculations for NeH and NeH$^{+}$ in its ground state were carried out using the MOLPRO quantum chemistry program suite~\cite{werner2022} at the Multi-Configurational Self-Consistent Field (MCSCF) - Multi-Reference Configuration Interaction (MRCI) level of theory with a complete active space (CAS) of 8,3,3,0 orbitals in the C$_{2v}$ point-group symmetry. The CAS corresponds to the complete valence active space extended to include the $n=2$ and $n=3$ orbitals of hydrogen. We started with the augmented correlation-consistent polarized valence triple zeta (aug-cc-pVTZ) basis sets implemented in MOLPRO, and we have extended the hydrogenic part of the basis by two $s$, three $p$, and one $d$ diffuse atomic orbitals (AO) with exponents from Ref.~\cite{baer1994}. To summarize, the ground state of NeH$^{+}$, X $^{1}\Sigma^{+}$, and the lowest 7 states of NeH, respectively X $^{2}\Sigma^{+}$, A $^{2}\Sigma^{+}$, B $^{2}\Pi$, C $^{2}\Sigma^{+}$, 4 $^{2}\Sigma^{+}$, 2 $^{2}\Pi$, and 5 $^{2}\Sigma^{+}$ have been calculated~(Fig.~\ref{fig:1}).

The NACs between states of the neutral were calculated numerically using the MOLPRO derivative couplings (DDR) procedure for the following expressions~\cite{guberman1994, brady2024}:

\begin{widetext}
\begin{equation}
A_{ij}(R)=\biggl \langle \psi^{\Gamma}_{i}(\{\Vec{r}\},R)\bigg|\frac{\partial}{\partial R}\bigg|\psi^{\Gamma}_{j}(\{\Vec{r}\},R)\biggr \rangle_{\{\Vec{r}\}}
\label{AR}
\end{equation}

\begin{equation}
B_{ij}(R)=\biggl \langle \psi^{\Gamma}_{i}(\{\Vec{r}\},R)\bigg|\frac{\partial^{2}}{\partial R^{2}}\bigg|\psi^{\Gamma}_{j}(\{\Vec{r}\},R)\biggr \rangle_{\{\Vec{r}\}}=\frac{\partial A_{ij}(R)}{\partial R} - A_{ij}^{2}(R)
\label{BR}
\end{equation}
\end{widetext}

\noindent Here $\Gamma$ stands for the symmetry of the molecular state, $i$ and $j$ denote different electronic states belonging to the same symmetry, and we focused exclusively on couplings that satisfy $i<j$. $R$ is the internuclear distance, \{$\vec{r}$\} denotes the complete set of electronic coordinates linked to the electrons of the neutral. The couplings $A(R)$ and $B(R)$ between states dissociating up to the Ne+H($n=3$) limit - X~$^{2}\Sigma^{+}$ with 4~$^{2}\Sigma^{+}$, A~$^{2}\Sigma^{+}$ with 4~$^{2}\Sigma^{+}$, C~$^{2}\Sigma^{+}$ with 4~$^{2}\Sigma^{+}$, X~$^{2}\Sigma^{+}$ with 5~$^{2}\Sigma^{+}$, A~$^{2}\Sigma^{+}$ with 5~$^{2}\Sigma^{+}$, C~$^{2}\Sigma^{+}$ with 5~$^{2}\Sigma^{+}$ for the $^{2}\Sigma^{+}$ symmetry, and B~$^{2}\Pi$ with 2~$^{2}\Pi$ for the $^{2}\Pi$ symmetry were produced~(Fig.~\ref{fig:2}). The couplings involving the highest excited state 5 $^{2}\Sigma^{+}$ are not shown as their magnitudes are negligible (approximately $10^{-8}$ $a_{0}^{-1}$).


For both electronic symmetries, $^2\Sigma^{+}$ and $^2\Pi$, the quantum defects $\mu^{\Lambda}_{\ell}(R)$ were extracted from the potential energy curves (PECs) of the highest-lying Rydberg-like states for which non-adiabatic couplings (NACs) remain non-negligible, using the Rydberg formula:

\begin{equation}
U^\Lambda_{\nu}(R) = U^{+}(R) - \frac{\mathcal{R}}{\left[n - \mu^{\Lambda}_{\ell}(R)\right]^2},
\end{equation}
where $\mathcal{R}$ 
is the Rydberg constant, $U^{+}(R)$ is the PEC of the ionic ground state X~$^1\Sigma^{+}$, and $U^\Lambda_{\nu}(R)$ is the PEC of a Rydberg state characterized by the quantum numbers $\nu = (n, \ell)$, with $n$ being the principal quantum number and $\ell$ the orbital angular momentum of the Rydberg electron. Specifically, the 4~$^2\Sigma^{+}$ (3$s$) and 2~$^2\Pi$ (3$p$) states were used for this purpose, as they are the highest members of their respective Rydberg series for which NACs with dissociative Rydberg states are still significant. In particular, the higher 5~$^2\Sigma^{+}$ (3$p$) state was not used, since its coupling with dissociative Rydberg states was found to be negligible. The extracted quantum defects $\mu^{\Lambda}_{\ell}(R)$ are associated with the electronic symmetry $\Lambda$ and the partial wave $\ell$~(Fig.~\ref{fig:2}).

Dissociative recombination in the absence of curve crossing has been investigated in a limited number of studies, notably by Guberman~\cite{guberman1994} in the case of HeH$^{+}$, and more recently in the work of Kalosi \textit{et al.}~\cite{kalosi2024} on ArH$^{+}$. In such cases, the neutral Rydberg state does not intersect the potential energy curve of a dissociative valence state. Instead, predissociation occurs through non-adiabatic couplings to an adiabatic state that is energetically open for dissociation. These couplings originate from terms in the nuclear kinetic energy operator that are neglected under the Born–Oppenheimer approximation.

If we denote by $\psi_d(\vec{r}, R)$ the electronic wavefunction associated with a dissociative channel and by $\chi_d(R)$, its corresponding nuclear vibrational wavefunction, and label the highest Rydberg-like electronic state of a given series as $\psi_\nu(\vec{r}, R)$, with $\chi_{v^+}(R)$ the vibrational wavefunction of the ion, then the electronic–nuclear coupling matrix element between these states is given by:

\begin{equation}
\mathcal{V}_{d,\nu v^{+}} = \left\langle \psi_d  \chi_d \middle| \hat{T}_N\middle| \psi_\nu \chi_{v^+} \right\rangle.
\label{Nuclearmatrix_density}
\end{equation}
The nuclear kinetic energy operator in the molecular frame is given by \cite{lefebvre2012}:

\begin{equation}
\hat{T}_N = -\frac{\hbar^2}{2M_r} \left[ \frac{\partial^2}{\partial R^2} + \frac{2}{R} \frac{\partial}{\partial R} \right] + \frac{\hbar^2}{2 M_r R^2} \hat{R}^{2}.
\end{equation}
Here, $M_r$ denotes the reduced mass of the electron+molecular cation system.
Since the rotational part $\hat{R} = \hat{J} - \hat{L} - \hat{S}$ yields off-diagonal elements that scale as $J(J+1)$, being responsible for the rotational effects, and are generally small compared to the radial terms \cite{leoni1972}, we restrict our attention to the radial kinetic operator. Using the product rules for differentiation:

\begin{align}
\frac{\partial^2(\psi \chi)}{\partial R^2} &= \chi \frac{\partial^2 \psi}{\partial R^2} + \psi \frac{\partial^2 \chi}{\partial R^2} + 2 \frac{\partial \psi}{\partial R} \frac{\partial \chi}{\partial R}, \\
\frac{2}{R} \frac{\partial}{\partial R}(\psi \chi) &= \frac{2}{R} \chi \frac{\partial \psi}{\partial R} + \frac{2}{R} \psi \frac{\partial \chi}{\partial R},
\end{align}
we derive the full matrix element:

\begin{equation}
\begin{split}
\mathcal{V}_{d,\nu v^{+}} = -\frac{\hbar^2}{2M_r} \Bigg[ &
\left\langle \chi_d \middle| \left\langle \psi_{d} \middle|\frac{\partial^{2}}{\partial R^{2}} \middle| \psi_{\nu} \right\rangle \middle| \chi_{v^+} \right\rangle \\
&+ 2\left\langle \chi_d \middle| \left\langle \psi_{d} \middle| \frac{\partial}{\partial R} \middle| \psi_{\nu} \right\rangle \middle| \chi_{v^+} \right\rangle \\
&+ \left\langle \chi_d \middle| \left\langle \psi_{d} \middle| \frac{2}{R} \frac{\partial}{\partial R} \middle| \psi_{\nu} \right\rangle \middle| \chi_{v^+} \right\rangle 
\Bigg].
\end{split}
\label{H_density_general}
\end{equation}
This expression introduces the NACs, which we write as:

\begin{align}
A_{d\nu}(R) &= \left\langle \psi_d \middle| \frac{\partial}{\partial R} \middle| \psi_\nu \right\rangle, \label{NactheoA} \\
B_{d\nu}(R) &= \left\langle \psi_d \middle| \frac{\partial^2}{\partial R^2} \middle| \psi_\nu \right\rangle. \label{Nactheob}
\end{align}These are related via:

\begin{equation}
\frac{\partial A(R)}{\partial R} = \left\langle \frac{\partial \psi_d}{\partial R} \middle| \frac{\partial \psi_\nu}{\partial R} \right\rangle + \left\langle \psi_d \middle| \frac{\partial^2}{\partial R^2} \middle| \psi_\nu \right\rangle.
\end{equation}Assuming that the cross-term can be approximated as $A_{d\nu}^2(R)$ \cite{brady2024}, we have:

\begin{equation}
B_{d\nu}(R) = \frac{\partial A_{d\nu}(R)}{\partial R} - A_{d\nu}^2(R).
\end{equation}Substituting these into Eq.~(\ref{H_density_general}), we obtain:

\begin{equation}
\begin{split}
\mathcal{V}_{d,\nu v^{+}} = 
- \frac{\hbar^2}{2M_r} \Bigg[ &
\left\langle \chi_d \middle| \left( B_{d\nu}(R) + 2 A_{d\nu}(R) \frac{\partial}{\partial R} \right) \middle| \chi_{v^+} \right\rangle \\
&+ \left\langle \chi_d \middle| \frac{2}{R} A_{d\nu}(R) \middle| \chi_{v^+} \right\rangle
\Bigg].
\end{split}
\label{H_density_compact_factored}
\end{equation}
In this expression, the term involving $(2/R)\,A_{d\nu}(R)$ is eliminated by expressing the vibrational wave functions as $\chi(R) = \zeta(R)/R$, where $\zeta(R)$ is normalized with respect to the infinitesimal 1D  volume element $dR$~\cite{lefebvre2012}. This transformation introduces a compensating term upon differentiation, which exactly cancels the $(2/R)\,A_{d\nu}(R)$ contribution. As a result, Eq.~(\ref{H_density_compact_factored}) retains only the derivative-dependent terms involving $A_{d\nu}(R)$ and $B_{d\nu}(R)$.

In order to account for the transition into the ionization continuum, a radial density of states $\rho(R)$ needs to be included in the expressions for the non-adiabatic couplings in Eqs.~(\ref{NactheoA}) and~(\ref{Nactheob}). The energy of a Rydberg-like state with effective principal quantum number $n^* = n - \mu_{\ell}^{\Lambda}(R)$, 
is given by:

\begin{equation}
E_{n^*} = -\frac{\mathcal{R}}{(n^*)^2}.
\end{equation}
The corresponding density of states is then obtained as:

\begin{equation}
\rho(R) = \frac{dn^*}{dE_{n^{*}}}  = \frac{(n^{*})^3}{2\,\mathcal{R}},
\end{equation}or in atomic units:

\begin{equation}
\rho(R) = (n^{*})^3.
\end{equation}
The normalization integral of the Rydberg states assuming non-energy dependent quantum defects, are equal to the density of the states, so the norm of the wave function is the square-root of the above expression~\cite{jungen2011}: 

\begin{equation}
\beta_\nu(R) = (n^{*})^{3/2}.
\label{eq:rhoR_atomic}
\end{equation}
We can now introduce the electronic couplings with the continuum as:

\begin{equation}
\tilde{A}_{d\nu}(R)=\left\langle \psi_d \middle| \frac{\partial}{\partial R} \middle| \psi_\nu \beta_\nu \right\rangle, \label{NactheoAtilde}
\end{equation}and:

\begin{equation}
\tilde{B}_{d\nu}(R)=\left\langle \psi_d \middle| \frac{\partial^2}{\partial R^2} \middle| \psi_\nu \beta_\nu \right\rangle. \label{Nactheobtilde}
\end{equation}Differentiating Eq.~(\ref{NactheoAtilde}), we obtain:

\begin{equation}
\tilde{A}_{d\nu}(R)=\beta_\nu(R)A_{d\nu}(R)+\left\langle\psi_d \middle| \frac{\partial \beta_\nu}{\partial R} \middle| \psi_\nu \right\rangle,
\end{equation}and differentiating Eq.~(\ref{Nactheobtilde}), we get:

\begin{equation}
\tilde{B}_{d\nu}(R)=\beta_\nu(R)B_{d\nu}(R) + \left\langle\psi_d \middle| \frac{\partial^{2} \beta_\nu}{\partial R^{2}} \middle| \psi_\nu \right\rangle + 2\left\langle\psi_d \middle| \frac{\partial \beta_\nu}{\partial R} \frac{\partial}{\partial R} \middle| \psi_\nu \right\rangle.
\end{equation}
Since the quantum defect \(\mu_\ell^{\Lambda}(R)\) varies only weakly with the internuclear distance, as one can see in Fig.~\ref{fig:2}, the terms involving the derivatives of $\beta_\nu(R)$ can be neglected compared to those of \(\beta_\nu(R)A_{d\nu}(R)\) and \(\beta_\nu(R)B_{d\nu}(R)\).

And finally, the electronic couplings between the dissociation channels and the ionization continuum in atomic units read:

\begin{equation}
\begin{split}
\mathcal{\tilde{V}}_{d,\nu v^{+}} = 
- \frac{1}{2M_r}\left\langle \chi_d \middle| \beta_\nu(R)\left(B_{d\nu}(R) + 2 A_{d\nu}(R) \frac{\partial}{\partial R} \right) \middle| \chi_{v^+} \right\rangle.
\end{split}
\label{H_density_compact_factored2}
\end{equation}

\noindent In multichannel quantum defect theory, these matrix elements form the core of the interaction matrix $\boldsymbol{\mathcal{V}}$.

The effectiveness of multichannel quantum defect theory for modeling electron–diatomic cation collisions has been demonstrated in numerous previous studies across a variety of diatomic species. Notable examples include H$_2^+$ and its isotopologues, for which rotational and isotope effects as well as excited core states have been explicitly taken into account over a wide range of collision energies, from low to moderately high~\cite{motapon2008,waffeutamo2011,chakrabarti2013,motapon2014,Epee2016}. Core-excited effects have also been investigated for molecular ions such as CH$^+$\cite{mezei2019}, SH$^+$\cite{Kashinski2017,boffelli2023} and N$_2^+$~\cite{little2014,abdoulanziz2021}.

Dissociative recombination - Eq.~(\ref{DRVT}) - involves \textit{ionization} channels, describing electron–molecular cation interactions, and \textit{dissociation} channels, corresponding to atom–atom scattering. The interplay between these channels leads to quantum interference between two competing pathways: a \textit{direct} mechanism, in which capture occurs into energetically open autoionizing (doubly excited) dissociative state, and/or the \textit{indirect} mechanism, in which the electron is first captured into a (mono excited) bound Rydberg state associated with closed channels, followed by predissociation.

As the key elements of this theoretical framework have been detailed in prior works~\cite{mezei2019}, we restrict ourselves to a brief outline of the main steps of the method.

The initial step of our methodology involves constructing the {\it interaction matrix} {\Vmat} - Eq.~(\ref{H_density_compact_factored2}), which serves as the driving force during the collision process. Its elements quantify the interconnections between the ionization and the dissociation channels.

The next step consists of construction of the short-range {\it reaction matrix} {\Kmat}.
This is achieved using a second-order perturbative solution of the Lippmann-Schwinger equation. The {\Kmat}-matrix is then diagonalized, its eigenvalues being related to the long-range phase-shifts of the eigenfunctions. This is followed by a frame transformation from the Born-Oppenheimer (short-range) representation, characterized by 
$N$, $v$, and $\Lambda$ quantum numbers, valid for small electron-ion and nucleus-nucleus distances, to the close-coupling (long-range) representation, characterized by $N^+$, $v^+$, $\Lambda^+$ (for the ion), and $l$ (orbital quantum number of the incident/Rydberg electron), valid for both large distances.
Applying a Cayley transformation~\cite{cayley1894collected} to the frame transformation  matrices makes  it possible to construct  the {\it generalized scattering matrix} 
{\Xmat}. The elimination of the closed channels~\cite{Seaton1983} is then performed, yielding the {\it physical scattering matrix}
{\Smat}:
\begin{equation}
\Smat = \Xmat_{oo}-\Xmat_{oc}\frac{1}{\Xmat_{cc}-\exp({\rm -i 2 \pi} \numat)} \Xmat_{co}\,,
\label{eq:elimination}
\end{equation}
which relies on the block matrices built not only for the open channels, {\Xmat$_{oo}$},  but also for the closed ones, {\Xmat$_{oc}$, \Xmat$_{co}$} and \Xmat$_{cc}$. The diagonal matrix {\numat} appearing in the denominator of Eq.~(\ref{eq:elimination}) incorporates the effective quantum numbers corresponding to the vibrational thresholds of the closed ionization channels at the given total energy of the system.

Finally, the global cross section for the dissociative recombination is:
 \begin{equation}
\begin{split}
\sigma _{diss \leftarrow v_{i}^{+}}  &=\frac{\pi}{4\varepsilon} \sum_{(\Lambda,sym)}\rho^{(\Lambda,sym)} \sum_{\ell,j}\mid S^{(\Lambda,sym)}_{d_{j},v_{i}^{+}\ell}\mid^2,\label{eqDR} 
\end{split}
\end{equation}		
\noindent 
where $\varepsilon$ is the electron collision energy, $\Lambda$ is the projection of the electronic orbital angular momentum on the molecular axis, $\rho^{(\Lambda,sym)}$ is the ratio between the multiplicities of the neutral system and of the ion,~$sym$ is referring to the inversion symmetry  - {\it gerade/ungerade} - and to the spin quantum number of the neutral system.

The current MQDT calculations neglect rotational effects, since NACs among the states having different molecular symmetry were not included in the calculations. For each symmetry considered, the focus is placed on the lowest dissociative states accessible at low incident electron energies, below 2.26~eV, which corresponds to the ionization threshold of the ionic ground state. The calculations were performed for the three lowest vibrational levels of the ion, namely $v_i^{+} = 0$, $1$, and $2$.

\section{Results and discussion}

\begin{figure}[t]
\centering
		\includegraphics[width=0.90\columnwidth]{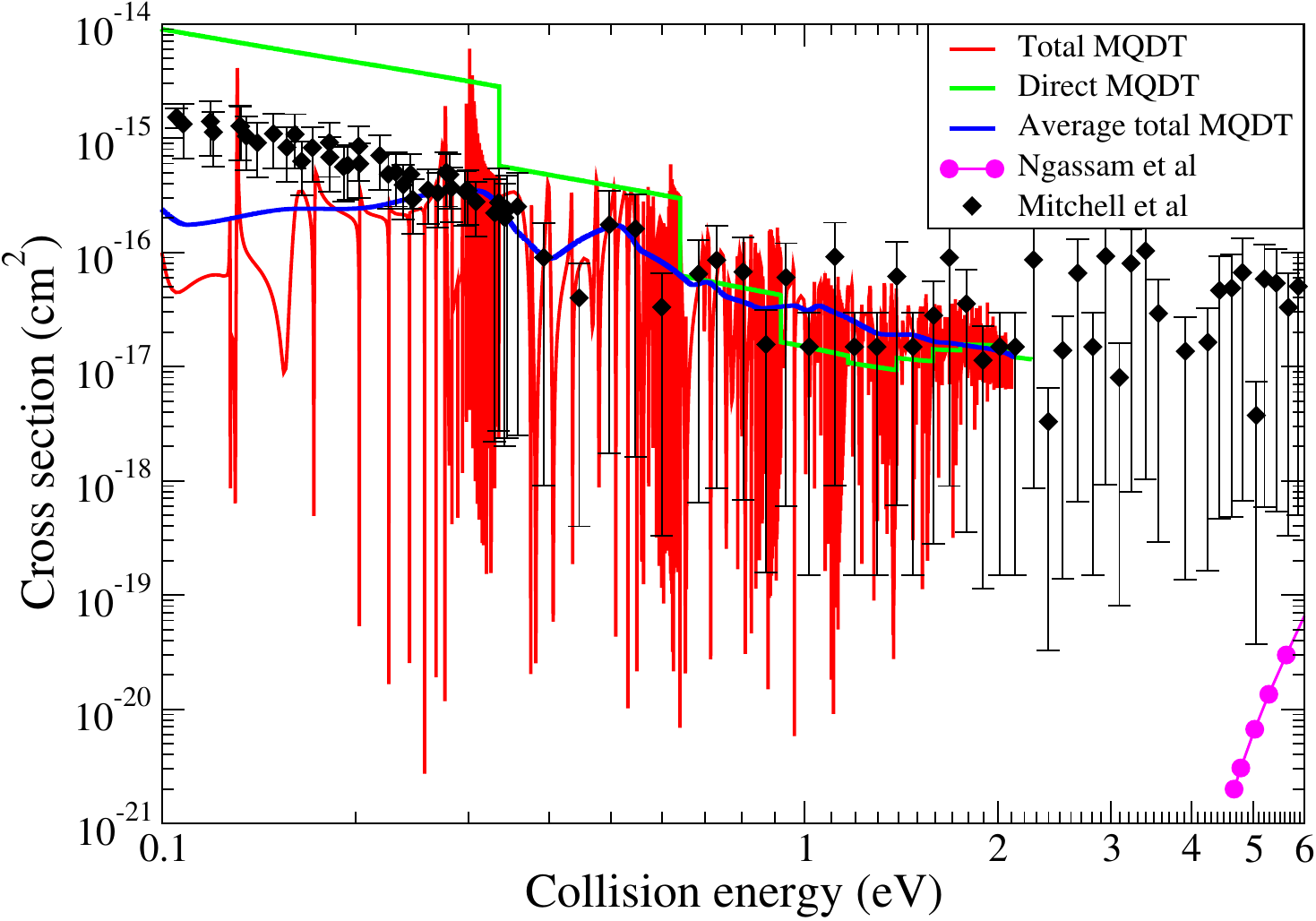}
    \caption{Dissociative recombination cross section of NeH$^+$ in its ground vibrational state ($v_i^+ = 0$), computed using MQDT with $^2\Sigma^+$ electronic states. The total and direct contributions are shown as red and green curves, respectively. The average total MQDT cross section is plotted in blue. Theoretical results from Ngassam et al.~\cite{ngassam2008} above 4.5 eV are shown in purple, and relative experimental measurements from the ASTRID storage ring by Mitchell et al.~\cite{mitchell2005} (black, with error bars) are shown for comparison, normalized to coincide with theory at 0.4 eV.}
    \label{fig:3}
\end{figure}
The contribution of the $^2\Pi$ symmetry to the dissociative recombination cross section is negligible compared to that of the $^2\Sigma^{+}$ symmetry. This is due to significantly weaker non-adiabatic couplings, which are found to be at least one order of magnitude smaller than those associated with the $^2\Sigma^{+}$ states, as one can see in Fig.~\ref{fig:2}. As a result, the global cross section is determined by the $^2\Sigma^{+}$ symmetry, and only this contribution will be shown in what follows.

Figure~\ref{fig:3} presents the dissociative recombination cross section of NeH$^+$ in its ground vibrational state ($v_i^+ = 0$), calculated using the upgraded MQDT. The results are shown together with the theoretical data of Ngassam et al.~\cite{ngassam2008} (magenta curves with full circles), calculated for collision energies above 4.5~eV, and the relative experimental cross sections of Mitchell et al.~\cite{mitchell2005} (black diamonds with error-bars), measured at the ASTRID storage ring. The relative experimental cross section was scaled to the theoretical prediction at 0.4 eV to enable direct comparison. Overall, we find good agreement with experiment: both the MQDT and ASTRID cross sections span the same magnitude range ($10^{-21}$--$10^{-15}$~cm$^2$) between 0 and 2.26~eV and exhibit similar energy dependence. The presence of sharp resonances (red curve in Fig.~\ref{fig:3}) reflects the indirect nature of the process and highlights the critical role of non-adiabatic couplings. The drops observed in the computed cross section, particularly visible in the direct cross section (green curve in the same figure), correspond to the sequential opening of ionization channels, which reduce the probability of dissociation by introducing competing electron escape pathways like autoionization, leading to vibrational excitations. The average total MQDT cross section (blue curve) was obtained by dividing the computed anisotropic rate coefficient by the relative electron–ion velocity, evaluated consistently with the collision energy. This latter cross section compares best with the experimental one, since they were obtained in the same way.

\begin{figure}[t]
\centering
		\includegraphics[width=0.90\columnwidth]{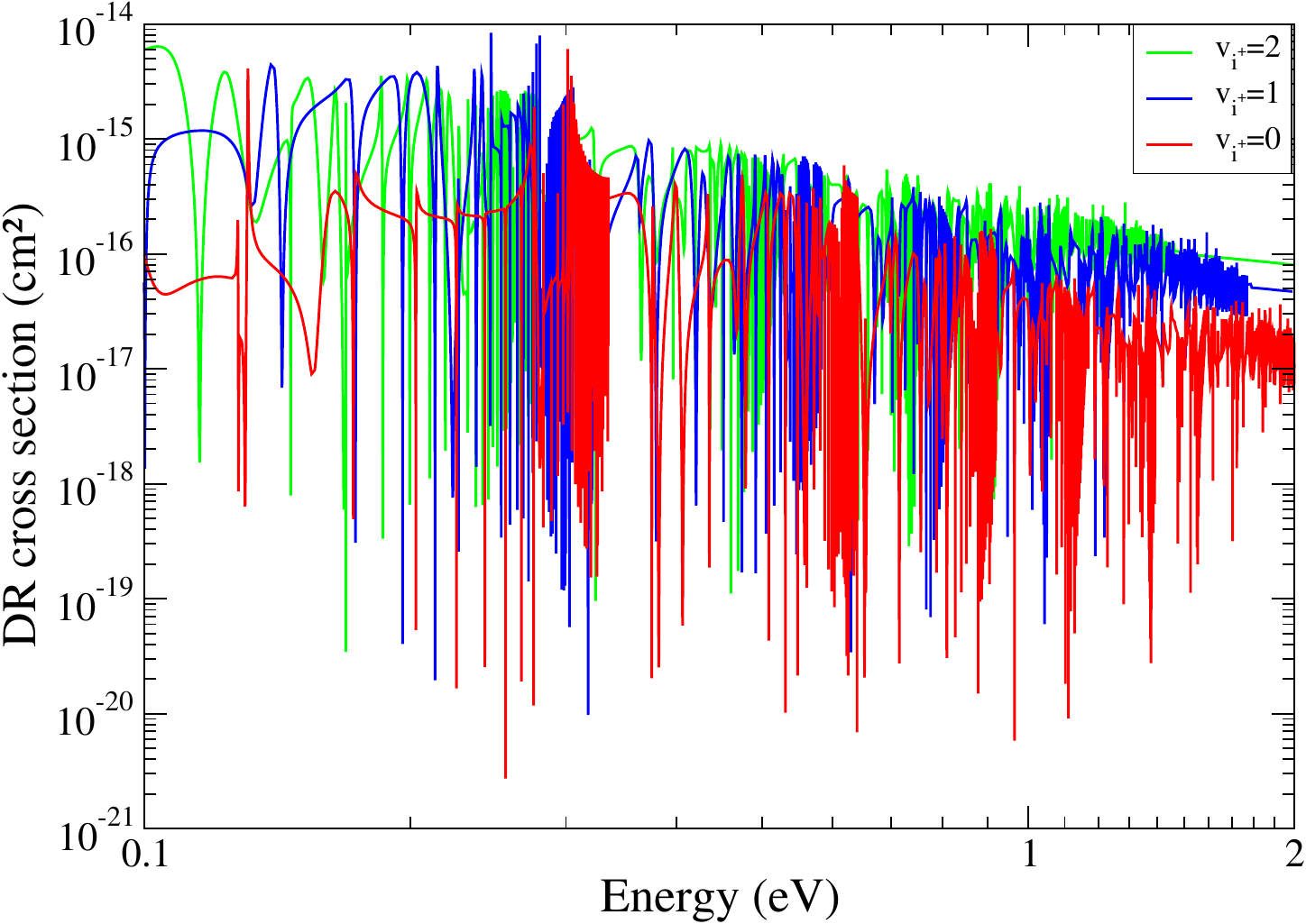}
    \caption{Dissociative recombination cross sections of NeH$^{+}$ for $v_i^+ = 0$, 1, and 2.}
    \label{fig:4}
\end{figure}

Figure~\ref{fig:4} presents the dissociative recombination cross sections for the first three vibrational levels of NeH$^{+}$. The slight increase observed from $v_i^+ = 0$ to $v_i^+ = 2$ can be attributed to the Franck–Condon overlap between the dissociative wave function and the derivative of the vibrational bound-state wavefunctions. As we increase the initial vibrational level of the cation target, the crossing among the two states becomes more favorable leading to higher direct cross section.

\begin{figure}[t]
\centering
	\includegraphics[width=0.90\columnwidth]{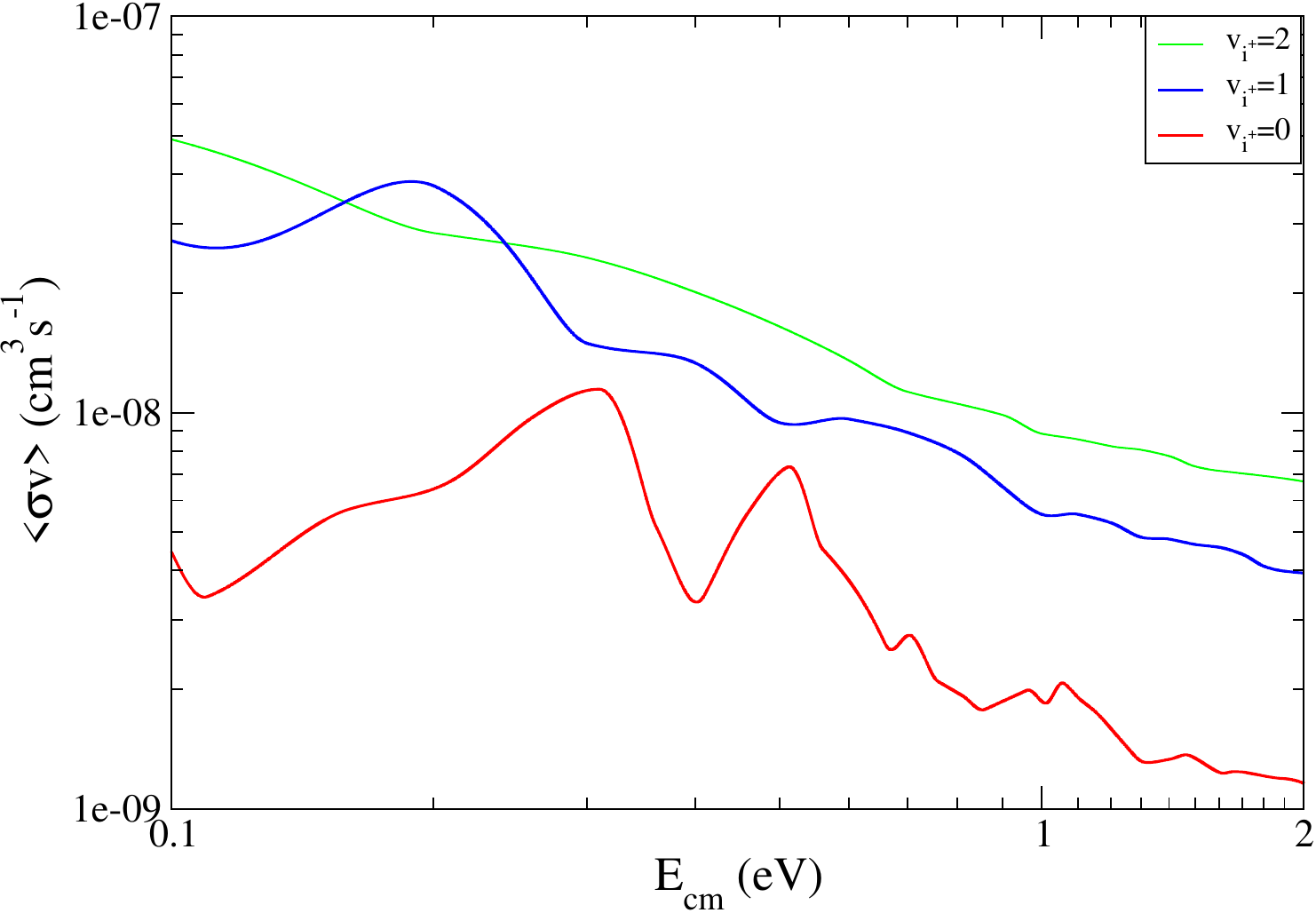}
    \caption{Anisotropic dissociative recombination rate coefficients of NeH$^+$ from its 3 lowest vibrational levels.}
    \label{fig:5}
\end{figure}

\begin{figure}
\centering
		\includegraphics[width=0.90\columnwidth]{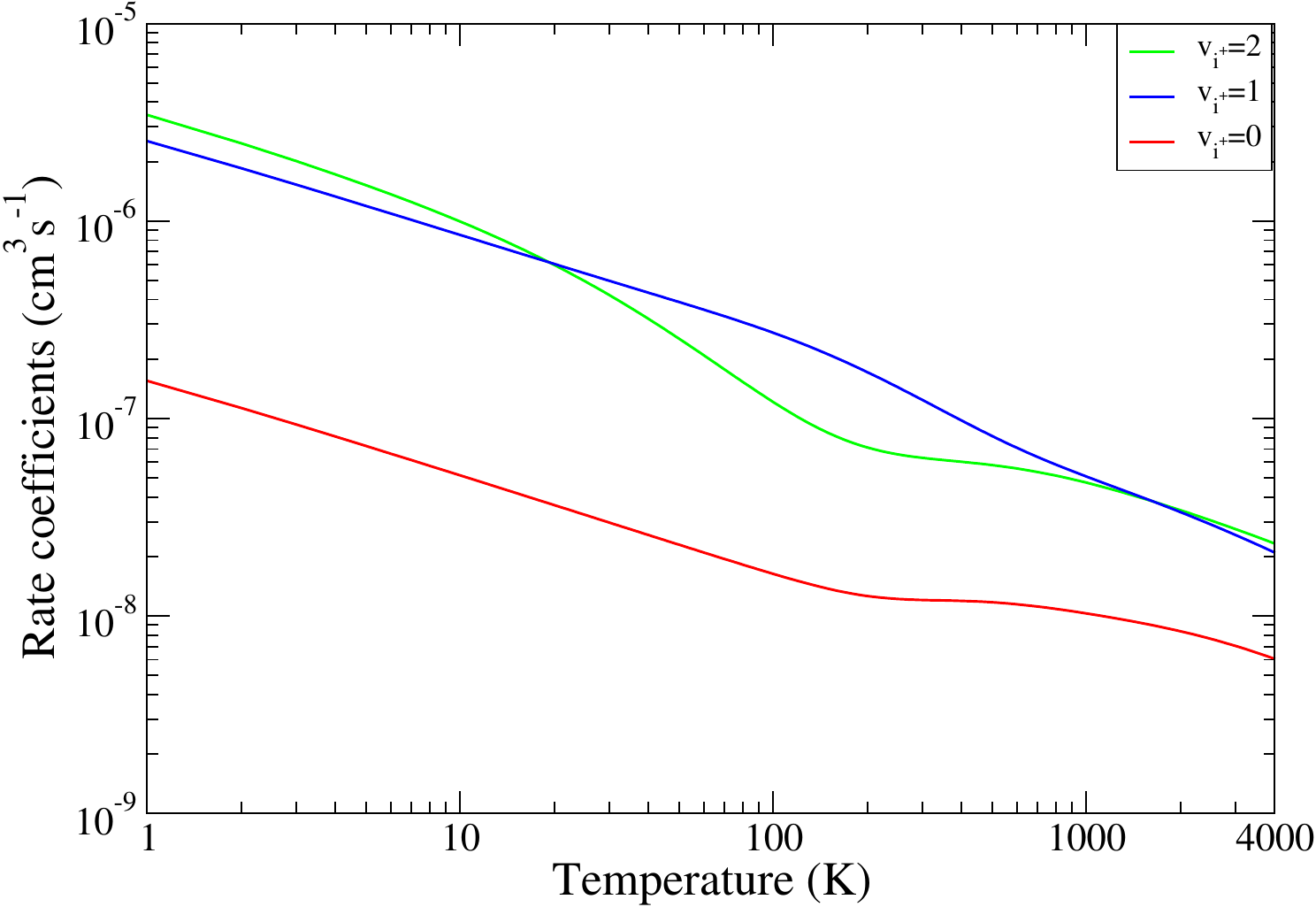}
    \caption{Isotropic dissociative recombination thermal rate coefficients of NeH$^+$ from its 3 lowest vibrational levels.}
    \label{fig:6}
\end{figure}
Figure~\ref{fig:5} shows convoluted cross sections or anisotropic rate coefficients for the DR of NeH$^+$ in its 3 lowest vibrational states ($v_i^+ = 0$, 1, and 2), calculated using an electron velocity distribution characterized by a longitudinal temperature of $kT_{\parallel} \approx 0.5~\text{meV}$ and a transverse temperature of $kT_{\perp} \approx 25~\text{meV}$ — parameters matching the experimental conditions reported by Mitchell et al.~\cite{mitchell2005}. The rate coefficients were obtained by convolving the MQDT cross sections with this anisotropic electron distribution: 
\begin{equation}
 f(v_d,{\bf v}) = \frac{m}{2\pi k T_\perp}exp{\left(-\frac{mv_\perp^2}{2k T_\perp}\right)}\sqrt{\frac{m}{2\pi k T_\parallel}}exp\left(-\frac{m\left(v_\parallel-v_d\right)^2}{2k T_\parallel}\right),
\end{equation}
where $v_d$, $v_\perp$, and $v_\parallel$ stand for the detuning (relative or collision), transversal, and longitudinal electron velocities. We obtain rates of the order of $10^{-9}$~cm$^{3}$\,s$^{-1}$ for $v_i^{+}=0$ ($4.4\times10^{-9}$ at 0.1~eV and $1.2\times10^{-9}$ at 2.0~eV), and about $10^{-8}$~cm$^{3}$\,s$^{-1}$ for $v_i^{+}=1$ ($2.7\times10^{-8}$ and $3.9\times10^{-9}$) and $v_i^{+}=2$ ($4.9\times10^{-8}$ and $6.7\times10^{-9}$).

In fusion edge plasmas containing neon, NeH$^+$ can form as a transient impurity through reactions with vibrationally excited H$_2^{+}$ ($v_{i}^{+} \geq 2$), which are expected to populate predominantly low vibrational levels $v_i^+ = 0$, 1, and 2 of NeH$^+$~\cite{ngassam2008,mitchell2005}. Our anisotropic DR rate coefficients for these vibrational levels are not negligible for molecular ions and show that dissociative recombination efficiently converts NeH$^+$ back into neutral neon and hydrogen atoms, preventing significant accumulation of the molecular and atomic ions in the divertor region.

Figure~\ref{fig:6} displays the isotropic thermal rate coefficients of DR calculated for $v_i^+ = 0$, 1, and 2, in the temperature range $1-4000$~K, relevant for stellar environments. The rate coefficients were derived from the cross sections by convolution with a Maxwell–Boltzmann isotropic electron energy distribution: 
\begin{equation}
 f({\bf v}) = \left(\frac{m}{2\pi k T}\right)^{3/2}exp{\left(-\frac{mv^2}{2k T}\right)}.   
\end{equation}
The values obtained span $5.2\times10^{-8}-6.1\times10^{-9}$~cm$^{3}$\,s$^{-1}$ for
$v_i^+=0$, $8.5\times10^{-7}-2.1\times10^{-8}$~cm$^{3}$\,s$^{-1}$ for $v_i^+=1$, and
$9.9\times10^{-7}-2.3\times10^{-8}$~cm$^{3}$\,s$^{-1}$ for $v_i^+=2$ over the temperature range $10-4000$~K.

In stellar remnants, the gas phase abundance of neon is substantially higher than that of argon
\cite{schwarz2002,Das2020}, which enhances the potential formation pathways of NeH$^{+}$ relative to ArH$^{+}$ under comparable conditions. In previous astrophysical simulations of nova and supernova remnants, the DR rate coefficients of ArH$^{+}$ and NeH$^{+}$ were both taken to be $\sim 10^{-11}$~cm$^{3}$\,s$^{-1}$, with the NeH$^{+}$ value assumed identical to that of ArH$^{+}$ \cite{Das2020,sil2024}. However, recent experimental measurements for ArH$^{+}$ in its ground vibrational state~\cite{kalosi2024} indicate thermal rates of $1.9\times10^{-9}$~cm$^{3}$\,s$^{-1}$ at $T=10$~K and $7.0\times10^{-11}$~cm$^{3}$\,s$^{-1}$ at $T=4000$~K, i.e.\ one to two orders of magnitude higher than the constant $10^{-11}$~cm$^{3}$\,s$^{-1}$ previously assumed in remnants models.

Our present DR calculations for NeH$^{+}$ yield substantially larger thermal rate coefficients for $v_{i}^{+}=0$, spanning $5.2\times10^{-8}$~cm$^{3}$\,s$^{-1}$ at $T=10$~K to $6.1\times10^{-9}$~cm$^{3}$\,s$^{-1}$ at $T=4000$~K, i.e.\ two to three orders of magnitude above the values previously adopted in remnants models. Such high rates imply that NeH$^{+}$ is destroyed efficiently, strongly reducing its equilibrium abundance and plausibly explaining its current non-detection. These updated DR coefficients should therefore be implemented in modeling frameworks such as \textsc{Cloudy} \cite{ferland1998}, which are widely applied to simulations of nova and supernova remnants.

\section{Conclusions}
In summary, we have performed MQDT calculations of the low-energy dissociative recombination of NeH$^{+}$, providing cross sections and rate coefficients resolved by initial vibrational level ($v_i^+=0-2$).

The present theoretical approach extends the treatment of non-adiabatic couplings within MQDT by explicitly including both the first-order terms $A(R)$ and the second-order terms $B(R)$. In addition, a radial density of states $\beta_\nu(R)$ is incorporated to describe the transition into the ionization continuum, which allows MQDT calculations to be applied up to collision energies of 2.26~eV.

The dissociative recombination process is dominated by $^2\Sigma^{+}$ symmetry states. For $v_i^+ = 0$, the computed dissociative recombination cross sections for NeH$^{+}$ range from $10^{-21}$ to $10^{-15}$~cm$^2$, confirming that the process is weak but not negligible at low collision energies. The agreement with available experimental data is satisfactory.

The dissociative recombination process exhibits increasing cross sections with higher vibrational levels. The $v_i^+ = 1$ and $v_i^+ = 2$ levels yield larger contributions than $v_i^+ = 0$, which can be attributed to improved Franck--Condon-type overlap between the dissociative wavefunctions and the vibrational wavefunctions of the ion.

In fusion edge plasmas containing neon, NeH$^{+}$ is produced mainly via reactions with vibrationally excited H$_2^{+}$ ($v_{i}^{+} \geq 2$) which are expected to populate predominantly low vibrational levels $v_i^{+} = 0-2$. The present anisotropic DR coefficients for these states ($10^{-9}$--$10^{-8}$~cm$^{3}$\,s$^{-1}$; e.g.\ for $v_i^{+}=0$, $4.4\times10^{-9}$ at 0.1~eV and $1.2\times10^{-9}$ at 2.0~eV; for $v_i^{+}=1$, $2.7\times10^{-8}$ and $3.9\times10^{-9}$; and for $v_i^{+}=2$, $4.9\times10^{-8}$ and $6.7\times10^{-9}$) are not negligible for molecular ions and indicate that DR efficiently converts NeH$^{+}$ back into neutral neon and hydrogen, thereby preventing significant accumulation in the divertor region.

In nova and supernova remnants, the calculated isotropic thermal DR coefficients of NeH$^{+}$
for $v_i^{+}=0$ span $5.2\times10^{-8}$~cm$^{3}$\,s$^{-1}$ at $T=10$~K to $6.1\times10^{-9}$~cm$^{3}$\,s$^{-1}$ at $T=4000$~K, i.e.\ two to three orders of magnitude
larger than the constant $10^{-11}$~cm$^{3}$\,s$^{-1}$ value previously adopted in remnants
models for both ArH$^{+}$ and NeH$^{+}$, and one to two orders of magnitude larger than the
latest experimental ArH$^{+}$ rates ( $1.9\times10^{-9}$~cm$^{3}$\,s$^{-1}$ at 10~K and
$7.0\times10^{-11}$~cm$^{3}$\,s$^{-1}$ at 4000~K ). These high rates imply efficient destruction
of NeH$^{+}$ and help explain its current non-detection.

Finally, we note that, as in all theoretical MQDT treatments, the accuracy of the present results is limited by the quality of the underlying \textit{ab initio} potentials and couplings and by the approximations inherent to the scattering model. A quantitative assessment of the uncertainty is difficult, but the overall level of agreement with the available experimental data for $v_i^+=0$ suggests that the present results provide a reliable basis for plasma and astrophysical applications.

\section*{Data availability}

The data supporting this article is available from the corresponding author.

\begin{acknowledgments}
The authors acknowledge support provided by the F\'ed\'eration de Recherche Fusion par Confinement Magn\'etique (CNRS and CEA), La R\'egion Normandie, LabEx EMC3 through projects PTOLEMEE,  COMUE Normandie Universit\'e and the Institute for Energy, Propulsion and Environment (FR-IEPE). J.Zs.M. is grateful for financial support from the National Research, Development and Innovation Fund of Hungary, under the FK 19 funding schemes with project number FK 132989. This work was granted access to the HPC/AI resources of [CINES/IDRIS/TGCC] under the allocation 2023-2024 [AD010805116R2] made by GENCI.
\end{acknowledgments}

\bibliography{apssamp}

\end{document}